\documentclass[twosided,showpacs,pre,preprint,amsmath,amssymb]{revtex4}
\usepackage{graphicx}
\usepackage{float}
\usepackage{flafter}
\usepackage{epsfig}%

\usepackage{color}%

\usepackage{bm}

\newcommand{\remove}[1]{{}}

\newcommand{\blue}[1]{{\textcolor{black}{#1}}}

\newcommand{\Tc}{T_{\text c}}
\newcommand{\Fe}{F_{\text e}}
\newcommand{\Qdot}{\dot{Q}}
\newcommand{\gammadot}{\dot{\gamma}}

\newcommand{\sigmay}{\sigma_{\text y}}
\newcommand{\sigmaystatic}{\sigma_{\mathrm{y}}^{\mathrm{static}}}
\newcommand{\sigmaydynamic}{\sigma_{\mathrm{y}}^{\mathrm{dynamic}}}

\newcommand{\tauLJ}{\tau_{\text LJ}}

\newcommand{\rhoA}{\rho_{\text A}}
\newcommand{\rhoB}{\rho_{\text B}}

\newcommand{\myeq}{\!=\!}

\newcommand{\tw}{t_{\text {w}}}
\newcommand{\kB}{k_{\text {B}}}

\newcommand{\fref}[1]{Fig.\ \ref{#1}}

\newcommand{\rcab}{r_{\text{c},\alpha\beta}}
\newcommand{\ekin}{e_{\mathrm{kin}}}
\newcommand{\gammamct}{\gamma_{\text{MCT}}}

\newcommand{\snapshot}[2]{
\unitlength=1mm
\begin{picture}(85, 45)(-#1, -#2)
\put(2,7){\epsfig{file=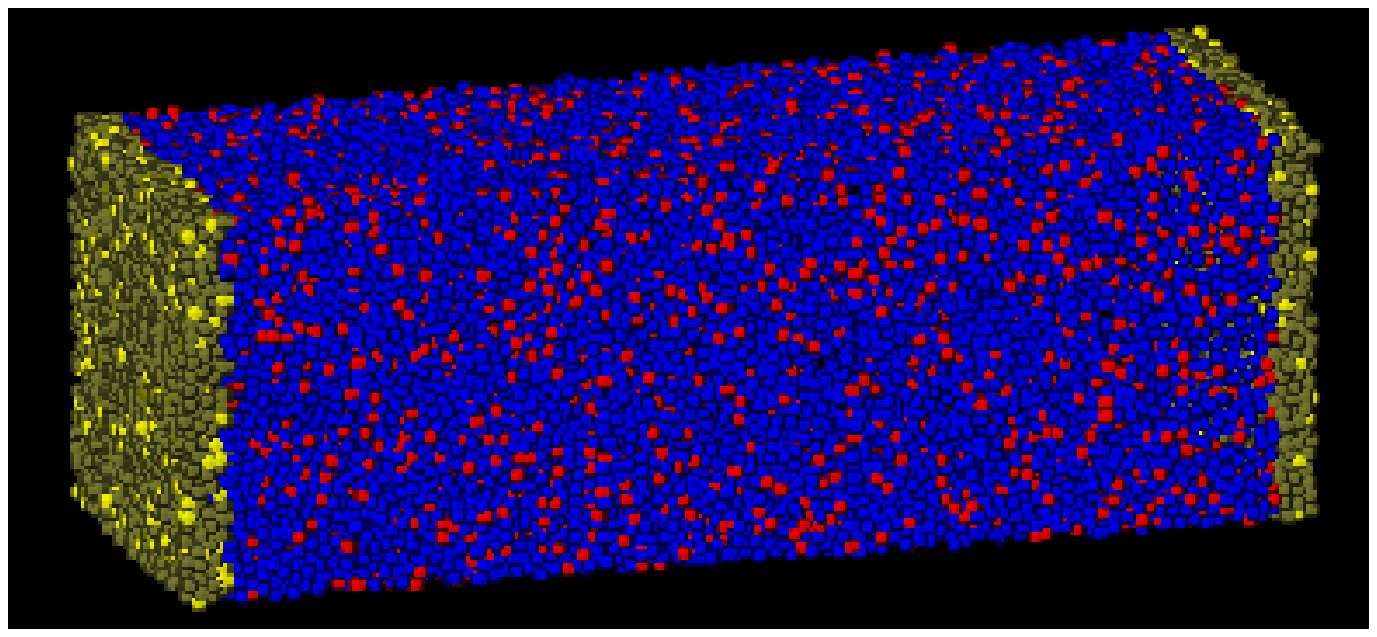,width=85mm,clip=}}
\thicklines
\put(-3,18){\large $x$}
\put(0,2){\line(0,1){20}}

\put(13,-1){\large $y$}
\put(0,2){\line(5,-2){10}}

\put(25,3){\large $z$}
\put(0,2.2){\line(5,0){30}}
\end{picture}
}

\begin{document}

\title{Profile blunting and flow blockage in a yield stress fluid:
A molecular dynamics study}

\author{F. Varnik and D. Raabe\\
\large{Max-Planck Institut f\"ur Eisenforschung}\\
\large{Max-Planck Stra{\ss}e 1, 40237 D\"usseldorf, Germany\\
Email: f.varnik@mpie.de}}

\date{\today}

\begin{abstract}
The flow of a simple glass forming system (a 80:20 binary
Lennard-Jones mixture) through a
planar channel is studied via molecular dynamics simulations. The
flow is driven by an external body force similar to gravity.
Previous studies show that the model exhibits both a static
[Varnik et al. J. Chem. Phys. 120, 2788 (2004)] and a dynamic 
[F. Varnik and O. Henrich Phys. Rev. B  73, 174209 (2006)] 
yield stress in
the glassy phase. \blue{These observations are corroborated by the
present work, where we investigate how the presence of a yield
stress may affect the system behavior in a Poiseuille-type flow
geometry.} In particular, we observe a blunted velocity profile
across the channel: A relatively wide region in the channel center
flows with a constant velocity (zero shear rate) followed by a non
linear change of the shear rate as the walls are approached. The
observed velocity gradients are compared to those obtained from the
knowledge of the shear stress across the channel and the flow-curves
(stress versus shear rate), the latter being determined in our
previous simulations of homogeneous shear flow.
Furthermore, using the value of the (dynamic) yield stress known
from previous simulations, we estimate the threshold body force for
a complete arrest of the flow. Indeed, a blockage is observed as the
imposed force falls below this threshold value. Small but finite
shear rates are observed at stresses above the dynamic but below the
static yield stress. We discuss the possible role of the
\blue{stick-slip like motion} for this observation.
\end{abstract}



\maketitle

\section{Introduction}
\label{sec:motivation}
The so called soft glassy materials
\cite{Sollich1997,Hebraud1998,Fielding2000}
exhibit a rich variety of interesting rheological phenomena.
When compared to a Newtonian liquid (defined as a liquid whose
viscosity, $\eta$, does not depend on shear rate, $\gammadot$)
the viscosity of a soft glassy system shows pronounced dependence on
imposed shear rate. To be specific, let us consider disordered colloidal
suspensions \cite{Laun1992,Petekidis2002,Petekidis2003} as a
typical example. It is well known that the shear viscosity
of dense colloidal dispersions decreases with increasing $\gammadot$
(shear thinning) if one focuses the attention on low shear rates.
In the limit of high shear rates, on the other hand, the shear viscosity
starts to increase with $\gammadot$ (shear thickenning).
While shear thinning is commonly attributed to the competition between
the time scale imposed by the external flow and the time scale of
inherent structural relaxation, shear thickenning phenomenon is
rather understood as originating from hydrodynamic effects \cite{Cates}
(whose contribution to the stress is negligible at low shear rates
and high concentrations but increases considerably at high $\gammadot$ \cite{Phung1996}).

\blue{In this paper, we study a model system whose rheological properties
can be rationalized without taking into account hydrodynamic effects
\cite{Varnik2006}.} Previous studies of the model
showed that it exhibits both a static and a dynamic yield stress.
The main difference between the static and the dynamic yield stress
lies upon the imposed quantity. While the static yield stress,
$\sigmaystatic$, is measured in experiments upon imposed stress,
the dynamic yield stress, $\sigmaydynamic$,
is measured in experiments upon imposed shear. This is so because a soft glassy material does
not develop the same rheological response in the both mentioned cases.

To see this, let us consider a planar Couette cell. If a lateral
force per unit area (=stress) is imposed to one of the walls of the
Couette cell,  the system may resist to the imposed stress if the
latter is below some threshold value (which generally depends on
temperature and density, say). For stresses higher than this
threshold value, on the other hand, the system is 'liquidized' and
flows with a linear velocity profile across the channel
\cite{Varnik2004}. The static yield stress thus characterizes the
response of an initially non-driven amorphous solid.

If instead of the stress an average shear rate is imposed
(by, e.g., moving one of the walls with a constant velocity), the occurrence of a flow is
unavoidable by construction. However, the flow profile need not be linear in
this case and may exhibit strong spatial heterogeneity.
In particular, a two-phase scenario may occur:
A region of zero shear ('solid-like' or 'jammed')
coexisting with a sheared ('liquid-like') region \cite{Varnik2003}.
Interestingly, as the wall velocity approaches zero, the shear stress across
the system does not converge to zero (as would be the case for a Newtonian
fluid) but seems to saturate at a finite value
\cite{Varnik2003,Fuchs2005,Varnik2006}. This limit of the shear stress
for vanishing shear rate is usually defined as dynamic yield stress.
It follows from the above description that the dynamic yield stress
characterizes the response of a shear molten amorphous solid.

It is noteworthy that the static yield stress is found to be higher
than its dynamic counterpart \cite{Varnik2004}, a situation
reminiscent of the difference between static and dynamic friction
(without any claim for a strict analogy). For the present model and
at a temperature and density of $T=0.2$ and $\rho=1.2$ our previous
studies yield $\sigmaydynamic \approx 0.5$ \cite{Varnik2006} and
$\sigmaystatic \approx 0.6$  \cite{Varnik2004} (all quantities are
given in Lennard-Jones (LJ) units; see below).

\blue{In a glass forming system, the static yield stress
does in general depend on the system history
and the way the stress is imposed. In particular, it depends
on the waiting time (the time elapsed between the temperature/density
quench and the beginning of stress ramp) as well as on the rate with which the
stress is increased. In the present simulations, however, no stress ramp is
applied. Rather, after a waiting time of $\tw$, we instantaneously
switch on a constant external force field exerting a body force of $\Fe$
on each particle (see also section \ref{sec:model}).
Furthermore, we focus on sufficiently
large waiting times so that, within the time window accessible to our
simulations, flow profiles are hardly affected by a dependence
of the static yield stress upon $\tw$
(see section \ref{sec:results} for further discussion of this issue).}

\blue{In contrast to the static yield stress, the above definition of a \emph{dynamic}
yield stress allows one to avoid these complications. This follows from the fact that
$\sigmaydynamic$ reflects the system response to a small but finite steady state shear,
where time translation invariance is recovered. Taking the limit of vanishing shear
rate does not affect this property.}

\blue{Obviously, the existence of a dynamic yield stress presupposes
the existence of a plateau in the flow curve (shear stress versus
shear rate) in the limit of low shear rates. However, although our
previous studies support the existence of such a plateau in the case
of the present binary LJ model, the issue of a dynamic yield stress
in soft glassy materials still remains controversial (see e.g.\
\cite{Besseling2007} for recent experimental studies of this topic
in colloidal dispersions). Therefore, it is worth to check whether
our model system also exhibits other features which follow from the
existence of a yield stress. As will be shown in this report, the
answer to this question is affirmative. In particular, we observe
non trivial behavior such as profile blunting and flow blockage in a
Poiseuille-type flow geometry, features which can consistently be described assuming
the existence of a dynamic yield stress.}

The paper is organized as follows. After an introduction of the
model and the simulation method in the next section, the effect of
yield stress on the behavior of the system in a planar channel flow
driven by an external body force will be investigated. In particular, it
will be shown that the velocity profile exhibits salient features of a
two-phase system: A 'solid-like' central part with zero velocity gradient
and two lateral 'liquid-like' sections between the channel center
and the walls. A consequence of this property on the dependence of the
mass flow rate upon the imposed force is worked out and compared to
the case of the same system in the normal liquid state, where it
behaves like a Newtonian liquid with a shear independent viscosity.
A summary compiles our results.

\section{A binary Lennard-Jones mixture}
\label{sec:model} In order to address the above mentioned issue, we
study via molecular dynamics simulations a generic glass forming
system, consisting of a 80:20 binary mixture of Lennard-Jones
particles (whose types we call A and B) at a total density of
$\rho\myeq \rhoA+\rhoB \myeq 1.2$.

A and B particles interact via
$U_{\text{LJ}}(r)\myeq
4\epsilon_{\alpha\beta}[(d_{\alpha\beta}/r)^{12}-(d_{\alpha\beta}/r)^6],$
with $\alpha,\beta\myeq {\text{A,B}}$, $\epsilon_{\text{AB}}\myeq
1.5\epsilon_{\text{AA}}$, $\epsilon_{\text{BB}}\myeq
0.5\epsilon_{\text{AA}}$, $d_{\text{AB}}\myeq 0.8d_{\text{AA}}$,
$d_{\text{BB}}\myeq 0.88d_{\text{AA}}$ and $m_{\text{B}}\myeq
m_{\text{A}}$. {\blue In order to enhance computational efficiency,}
the potential was truncated at twice the minimum
position of the LJ potential, $\rcab\myeq 2.245 d_{\alpha\beta}$.

\blue{The truncation of the LJ potential introduces a discontinuity in the force
field, which could be corrected via a smoothing procedure \cite{Haile1992}. However, the
present model with the truncated version of the LJ potential
has extensively been studied in the literature and has become a benchmark model
for the studies of glassy systems. Therefore, and for the purpose of comparison with
previous studies \cite{Kob1994,Kob1995,Kob1995a,Nauroth1997,Berthier2002a,Varnik2003,Varnik2004},
 we keep the model as it is. Note also that the use of a truncated
LJ potential is not \emph{a priori} a disadvantage, since we do not
seek a comparison with analytic studies of this specific system.
Rather, we are interested in generic features of a glass forming
model system for which the present binary LJ mixture has indeed
become a prototypical example.}

The parameters
$\epsilon_{\text{AA}}$, $d_{\text{AA}}$ and $m_{\text{A}}$ define
the units of energy, length and mass.  All other quantities reported
in this paper are expressed as a combination of these units.
The unit of time, for example, is given by
$\tauLJ \myeq d_{\text{AA}}\sqrt{m_{\text{A}} / \epsilon_{\text{AA}}}$ and that of stress
by $\epsilon_{\text{AA}} / d^3_{\text{AA}}$. Equations of
motion are integrated using a discrete time step of $dt \myeq 0.005$.
\blue{In order to test numerical accuracy, we also performed simulation runs using
a smaller time step of $dt\myeq 0.001$. Since no deviations were found between the
results obtained  for $dt\myeq 0.005$ and $dt\myeq 0.001$, we chose the larger
time step for all the subsequent simulations.}

\begin{figure}
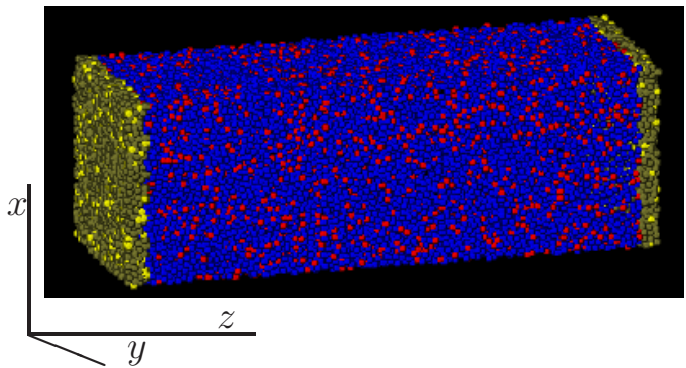

\snapshot{0}{0} \caption[]{(Color online) A snapshot of the
simulation box. The system consists of 80 percent of particles of
type A (blue) and 20\% of type B (red). Two atomistic walls
(yellowish colors) confine the system along the $z$-direction (to
the left and right of the image). Periodic boundary conditions are
applied in the remaining $x$ and $y$ directions. The system size is
$ L_x \times L_y \times L_z = 30 \times 30 \times 86$. It contains
92880 particles in total. \blue{The center of the coordinate system
($\vec{r}=\vec{0}$) is placed in the middle of the simulation box.
Thus, $x\in [-L_x/2\;\; L_x/2]$, $y\in [-L_y/2\;\; L_y/2]$ and $z
\in [-L_z/2\;\; L_z/2]$. $x$ is the flow direction, $y$ the neutral
(vorticity) direction and $z$ the direction of the velocity
gradient. All lengths are in LJ units.} }
\end{figure}

The present model has first been introduced by Kob and Andersen in the context of the
dynamics of supercooled liquids \cite{Kob1994,Kob1995,Kob1995a}, who showed
that it was suitable for an analysis
of many aspects of the mode coupling theory of the glass transition (MCT)
\cite{Bengtzelius1984,Gotze1989,Gotze1992}.
In particular, at a total density of $\rho \myeq 1.2$, equilibrium studies of the model showed
that the growth of the structural relaxation times at low
temperatures could be approximately described by a power law as predicted by the
ideal MCT, $\tau_{\text{relax}}\propto (T-\Tc)^{-\gammamct}$. Here,
$\Tc \myeq 0.435$ is the mode coupling critical temperature of the model
and $\gammamct$ is the critical exponent. For the present binary
Lennard-Jones system, numerical solution of ideal MCT equations
yields a value of $\gammamct \approx 2.5$ \cite{Nauroth1997}. A
similar value is also obtained for a binary mixture of soft spheres
\cite{Barrat1990}.

The model has also been studied in the context of the
rheology of disordered systems \cite{Berthier2002a,Varnik2003,Varnik2004,Varnik2006,Varnik2006b}.
In these studies, aspects such as flow heterogeneity \cite{Varnik2003},
structural relaxation under external drive \cite{Berthier2002a}
and the existence of a static \cite{Varnik2004} and a dynamic \cite{Varnik2006}
yield stress were addressed.

An interesting consequence of the presence of a yield stress on the flow
behavior of the system is presented here. For this purpose, we simulate
a Poiseuille-type flow in a planar channel. The study of such a situation
is interesting since the stress in a Poiseuille-type flow is zero
in the channel center and increases linearly with the distance from it.
As will be shown below, this is a consequence of  momentum balance equation
\cite{Todd1995,Varnik2000} and thus independent of the specific flow profile
formed across the channel. If the fluid under consideration exhibits
a finite yield stress, one may expect that the fluid portion in a
certain region around the channel center (where the stress is below yield stress)
should behave like a solid body while it should flow
like a liquid further away from this region.

We first equilibrate a system of size $L_x\times L_y\times L_z=30
\times 30 \times 86$ (containing 92880 particles) at a temperature
of $T=0.45 (>\Tc)$. \blue{The temperature is then set to $T=0.2$
($<\Tc$) corresponding to a glassy phase. Note that, the velocity
distribution adapts itself very fast (within a time of one LJ unit)
to the Maxwell distribution corresponding to the new temperature.
Particle configurations, on the other hand, keep the memory of the
old (high) temperature for far larger times (see e.g.\
\cite{Varnik2004} for a short discussion of aging in the case of the
present model system).}

\blue{However, as time proceeds, particles gradually rearrange and the system
moves towards that part of the configurational space which corresponds to
the new (low) temperature. This process is fast in the beginning, where
thermodynamic driving forces are the largest, but slows down with time.
In fact, as a characteristic feature of a glassy phase, the system never
reaches thermal equilibrium. Rather, it keeps evolving towards it endlessly (aging).}

\blue{As a result of this aging process, time translation invariance is violated
and those system properties which would be independent on the measurement time
in an equilibrium state (such as structural relaxation time, diffusion coefficient,
static structure function, etc.) show a dependence on the time
elapsed between the temperature quench and the beginning of the measurement.
Furthermore, quantities computed as time averages also show a dependence on the
\emph{duration} of the measurement, thus reflecting the fact that, in an aging system,
ensemble and time averages are no longer equivalent.}

\blue{In particular, in a glassy state, inherent system dynamics slows
down upon aging and structural relaxation times grow (ideally) endlessly,
eventually exceeding any other time scale in the problem (such as the time
scale imposed by external shear). In this interesting limit, the system no longer
behaves like a liquid but rather exhibits solid-like properties, as
exemplified by the presence of a finite static yield stress.
Since we wish to concentrate on this late time behavior, we first run
simulations in the quiescent state for a sufficiently large waiting
time of $\tw=10^4$ LJ units before imposing an external force.}

\blue{A time of $\tw=10^4$ LJ units in our simulations is sufficiently large in the following sense.
First, it is large so that the system exhibits a finite, measurable static yield stress \cite{Varnik2004}.
Second, it is large enough so that the increase of the static yield stress upon further
aging is slow and does not lead to a qualitative change in the flow behavior
(see section \ref{sec:results} for a more detailed discussion of aging
effects on the flow behavior).}

After this initial period of time, two solid walls are introduced
parallel to the $xy$-plane by immobilizing all particles whose
$z$-coordinate satisfies $|z|>40$ (walls of three particle diameter thickness).
A flow is then imposed along the $x$-axis by applying on each particle a
constant force, $\Fe$. This gives rise to a force density
of $f=\rho \Fe$. \blue{The force on a particle is the
\emph{sum} of the interaction forces arising from the Lennard-Jones potential and $\Fe$ 
(recall that $\Fe=0$ for $t<\tw$).}

For our geometry with planar solid walls, it follows that
$(\vec{u} \cdot \vec{\nabla})\vec{u} = \vec{0} $, where $\vec{u}=(u(z),0,0)$ is the streaming velocity
(\blue{recall that $x$ is the flow direction,
$y$ the neutral  (vorticity) direction and $z$ the direction of the velocity gradient}).
In the steady state and in the absence of
a pressure gradient ($\vec{\nabla}p=\vec{0}$), the momentum
continuity equation thus reduces to $\partial{\sigma}/\partial{z}=f=\rho\Fe$,
which yields $\sigma(z)=\rho \Fe z$, where we used the symmetry of the shear
stress with respect to the $xy$-plane ($\sigma(z=0)=0$).

The external force does work on the system. This work is transformed
into heat via viscous dissipation. In the absence of a thermostat,
this would lead to a continuous increase of temperature with time.
In order to keep the system temperature at a prescribed value, the
viscous heat must be removed. For this purpose, we divide the system
into parallel layers of thickness $dz=1$ and rescale (once every 10
integration steps) the $y$-component of the particle velocities
within the layer, so as to  impose the desired temperature $T$. More
precisely, we first compute the local kinetic energy per particle
$\ekin=1/N(z)\sum m_i v_{yi}^2$ within a layer centered at $z$.
Here, $m_i$ is the mass of a particle, $N(z)$ the number of
particles in the layer and the sum runs over the particles in the
layer only. A scale factor, $s$, is then determined via the
requirement that the new velocities $sv_{y,i}$ satisfy $1/N(z)\sum
m_i (sv_{yi})^2 = \kB T$. This gives $s=(\kB T/\ekin)^{1/2}$.
Finally, the new velocities are computed via multiplication of
$v_{y,i}$ with $s$.

Note that such a local treatment is necessary to keep a homogeneous
temperature profile when the velocity profile is not linear.
This is so because in this case the shear rate, $\gammadot$,
and hence the rate of heat production, $\eta\gammadot^2$ may vary
within the channel giving rise to a temperature gradient if only
the average temperature in the channel is adjusted.

However, despite this local temperature control, the heat production
close to the walls (where the shear rate is very high)
is so fast that a temperature increase in this part of the channel can not be fully
avoided (recall that viscous heat scales with $\gammadot^2$).
The magnitude of the excess temperature strongly depends
on the applied force per particle and is practically negligible
for $\Fe<0.02$ (see also the discussion of \fref{fig:Tprofile}).

\section{Results}
\label{sec:results} Let us first examine possible effects of aging  
\blue{and flow time} on the velocity profile. For this purpose, we prepare the system as
described above for various choices of the waiting time $\tw$.
Recall that $\tw$ is the time interval between temperature quench
and the time at which the external force is switched on. At this
instant, we set the clock to zero, $t=0$ thus marking the onset of
body force.

\blue{While the system would gradually 'solidify' in the quiescent state,
the external force tries to induce a flow and thus tends to 'fluidize' (rejuvenate)
the system. The extent and the rate of this rejuvenation does, however, strongly
depends on the magnitude of the stress formed across the channel.
As mentioned above, the stress in the present Poiseuille-type geometry
is a linear function of the distance from the channel center: $\sigma(z)=\rho \Fe z$.
For a given external force, the shear stress thus increases when approaching the walls.}

\begin{figure}
\epsfig{file=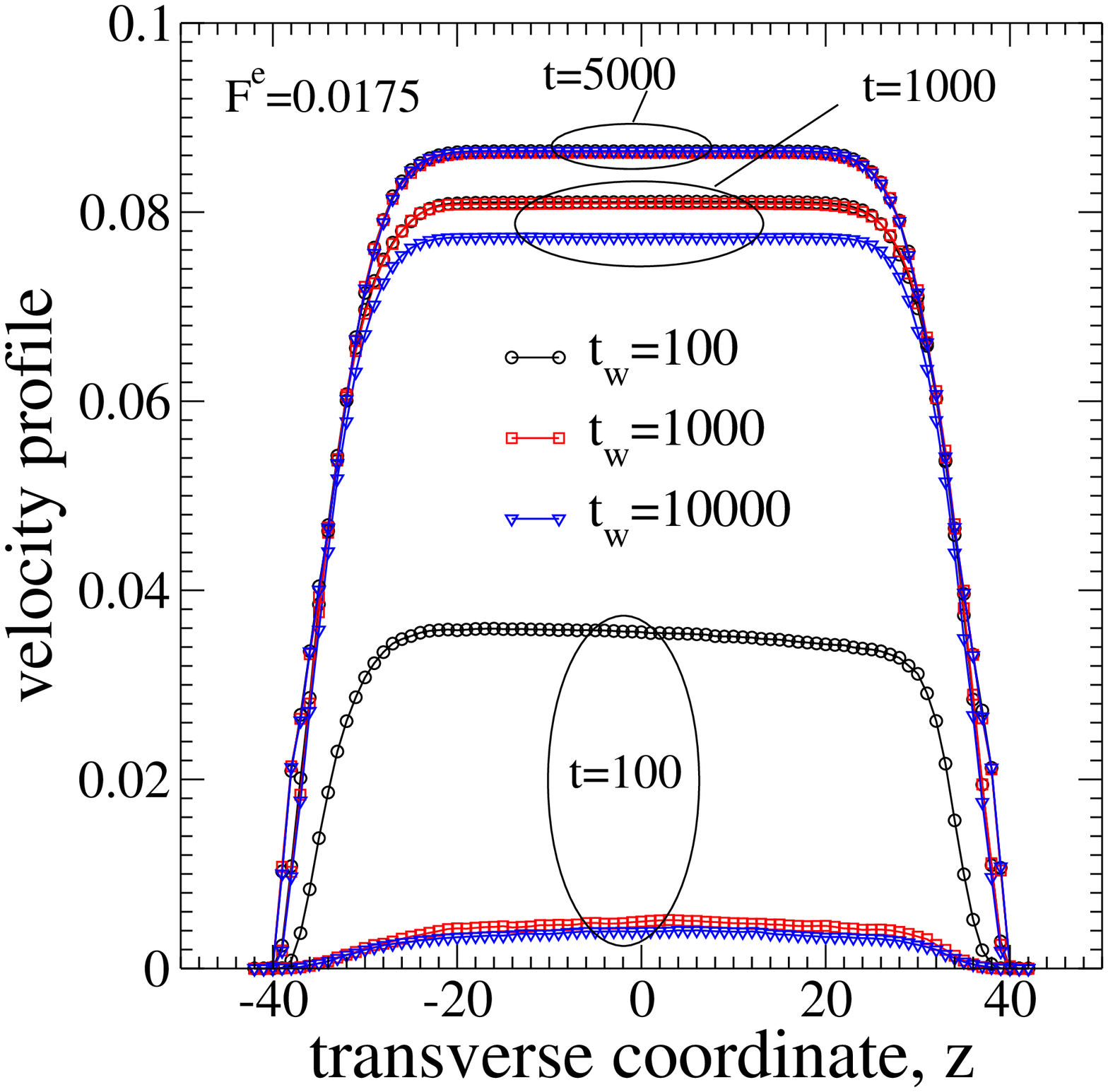,height=65mm,clip=}\hspace*{5mm}
\epsfig{file=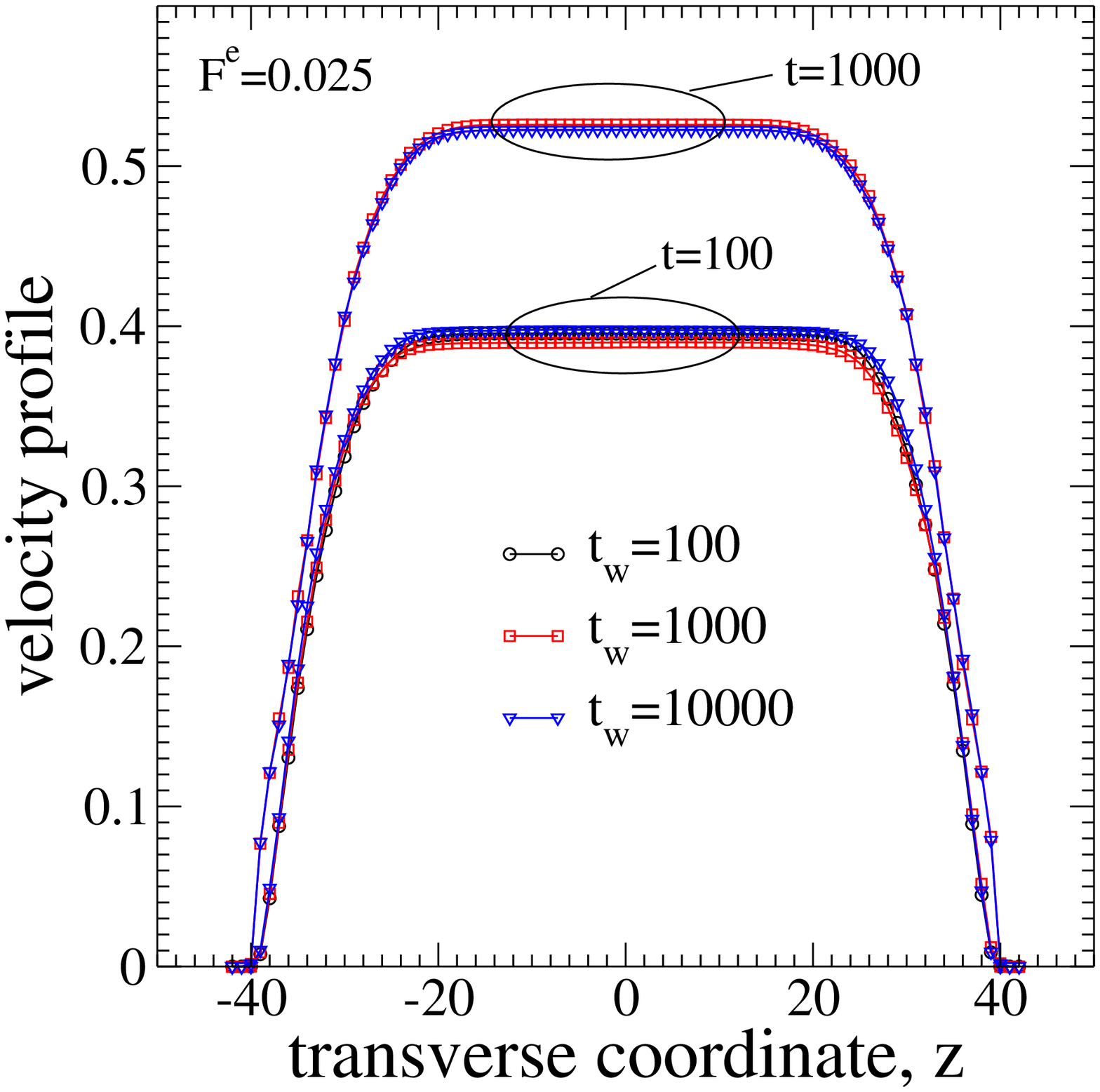,height=65mm,clip=} \caption[]{(Color online)
Competitive effects of aging and shear rejuvenation on the velocity
profile. Left: Velocity profile for waiting times of $\tw=10^2,\;
10^3$ and $10^4$ (LJ units) and various measurement times, $t$, as
indicated. Here, $\tw$ is the time elapsed between the temperature
quench and the onset of forcing ($\Fe=0.0175$). Data collection
starts at times $t=10^2,\; 10^3$ and $5\times 10^3$ after switching
on the external field. In order to better reveal the effects of
aging, the \emph{duration} of a measurement is limited to
$dt=10^2,\; 10^3$ and $5\times 10^3$, respectively. Right: Similar
plot as in the left panel, but for a higher applied force of
$\Fe=0.025$. Here, data collection starts at times $t=10^2$ and
$10^3$ after switching on the external field and spans over
$dt=10^2$ and $10^3$, respectively. Due to stronger forcing, effect
of shear rejuvenation shows up at shorter times compared to the left
panel. Note that the system response is independent of $\tw$ even at
$t=10^2$ (the observed difference between the curves is statistical,
otherwise a \emph{smaller} deformation would be observed at
$\tw=10^4$ compared to $\tw=10^3$). All quantities are in LJ units.
} \label{fig:vprofile-various-tw}
\end{figure}

\blue{As a consequence, the system flows first in the proximity of the walls,
while the center of the channel behaves rather like a solid body.
The effect of aging now reflects itself in the onset of the flow.
The larger $\tw$ the slower the initial flow behavior. This feature is
nicely born out in the left panel of \fref{fig:vprofile-various-tw}.
As a survey of the data corresponding to $t=10^2$ reveals, the system hardly
flows for $\tw=10^3$ and $\tw=10^4$ (deforming rather like an elastic solid)
 while for $\tw=10^2$ it 'liquidizes' in the proximity of the walls.}

\blue{The data shown in the left panel of \fref{fig:vprofile-various-tw} also
suggest that, for larger waiting times, the external force must be applied
during a longer period of time (larger $t$ in the figure) in order to remove
memory effects. This is seen by a closer look at the data corresponding to $t=10^3$.
Here, the curves belonging to $\tw=10^2$ and $\tw=10^3$ are practically identical,
while that of $\tw=10^4$ shows significantly slower deformation behavior.
However, applying the external force during a time of $t=5\times 10^3$, aging effects
disappear also in the case of $\tw=10^4$.}

\blue{For a given distance from the channel center, the shear stress increases (and hence
the time scale imposed by the external force decreases) with increasing the magnitude
of the applied force. This is an important aspect since memory effects (such as effect
of aging) decay within a typical structural relaxation time,
the latter being of the order of the externally imposed time scale
in glassy systems (see e.g.\ \cite{Varnik2006b} for a detailed
discussion of this issue). In fact, aging effects are expected to be observable
as long as this externally imposed time scale is significantly larger
than the waiting time. Since this time scale reduces upon increasing
$\Fe$, aging effects are expected to also disappear faster.
This property is nicely born out in the right panel of
\fref{fig:vprofile-various-tw}, where data similar to the left panel
are shown for the choice of a larger external force per particle, $\Fe=0.025$.
As seen from this panel, already at a time of $t=10^2$, the flow behavior
of the system is practically independent of $\tw$, even for a waiting time of $\tw=10^4\gg t=10^2$.}

Unless otherwise stated, the waiting time is $\tw=10^4$ for all the
results presented below. After the onset of external force, we wait
another period of time (of duration $t=5\times 10^3$) before
starting the data collection. This helps to reduce the above
discussed transient effects related to a competition of aging and
shear. Note, however, that in a Poiseuille-type flow of a glassy
system, transient effects will always be present to some extent.
This is closely related to the fact that the shear rate approaches
zero as one moves from the walls toward the center of the channel,
thus giving rise to progressively large relaxation times. As a
consequence, the time necessary to establish steady state ideally
diverges in the center of the channel. This behavior is enhanced in
a yield stress fluid, where the zero-shear zone has a finite
extension comprising a region around the channel center where the
local shear stress is below the fluid's yield stress. The
'jammed' region corresponds  to this part of the system.

\begin{figure}
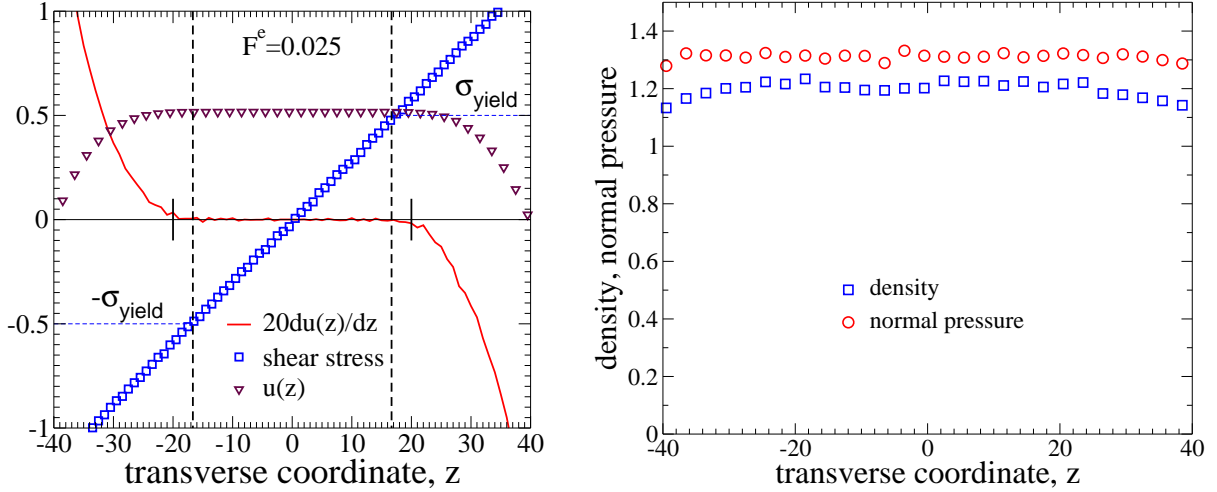

\epsfig{file=fig3a.ps,height=65mm,clip=}\hspace*{5mm}
\epsfig{file=fig3b.ps,height=65mm,clip=}
\caption[]{(Color online) Left: Flow through a three dimensional planar channel
of the model fluid studied in this paper.
The flow is generated by imposing
a constant body force of $\Fe=0.025$ on each particle. At a
temperature of $T=0.2$ and a total density of $\rho=1.2$, the model
exhibits a stress plateau at low shear rates which, for the present purpose,
plays the role of a dynamic yield stress, $\sigmaydynamic \approx 0.5$
\cite{Varnik2006}. One can easily show that the stress in
the channel behaves as $\sigma=\rho \Fe z$, where $z$ is the
transverse coordinate ($z=0$ being the channel center)
\cite{Varnik2002}. Obviously, for $ |z| \le 16.67 ( \approx \sigmay/(\rho\Fe))$,
the stress in the channel is below the dynamic yield stress. One thus
expects the system to behave as a solid body for $|z| \le 16.67$: Either it
should be at rest or move with a constant velocity (zero velocity
gradient). Indeed, an inspection of the velocity profile, $u(z)$, and its
derivative (rescaled to fit into the figure) confirms this
expectation (the region delimited by the two vertical dashed lines).
For $|z|>16.67$, on the other hand, the local shear stress exceeds
$\sigmaydynamic \approx 0.5$ leading to a liquid like
behavior. The system flows with a shear rate which non-linearly
increases upon increasing stress. The short vertical lines show the
limits of the expected 'jammed' region using the static yield stress,
$\sigmaystatic\approx 0.6$ (see also the text) \cite{Varnik2006b}.
Right: Profiles of the system density and the component of the
pressure perpendicular to the walls. Both the
density and the normal pressure are practically constant across
the channel. In particular, there is no wall migration in the case of
present simulations. This strongly suggests that the blunting of the
velocity profile observed in the left panel is related to the presence
of a yield stress which leads to the mentioned two-phase behavior.
All quantities are in reduced (LJ) units.}
\label{fig:yieldstressLJ}
\end{figure}

In the case of a Newtonian liquid, the above mentioned approach to drive the flow
would give rise to a parabolic velocity
profile of the form $u(z)=\rho\Fe(L_z^2/4-z^2)/(2\eta)$.
However, as Figs.\ \ref{fig:vprofile-various-tw} and \ref{fig:yieldstressLJ} clearly
demonstrate, the velocity profile in the case of the present model in the
glassy phase exhibits a quite different behavior:
In the central region, the velocity profile is flat with a zero gradient
while it gradually departs from this constant behavior (shear rate becoming non zero)
beyond this central part.

It is interesting to note that similar (blunted) shapes of the velocity
profile are also observed in pressure driven flows of both
neutrally buoyant suspensions of spheres (with a size of the order of 1mm)
\cite{Hampton1997,Han1999} as well as red blood cells
(biconcave disks of $2\mu$m thickness and $8\mu$m diameter)
\cite{Tangelder1986,Bishop2001a}. In these cases, however, the profile
blunting is usually accompanied by a migration of particles from the
wall region towards the center of the channel ('wall migration') a phenomenon,
which is absent in the case of present studies (see the right panel
of \fref{fig:yieldstressLJ}).

\blue{Other examples of blunted velocity profiles occur in systems with a
nematic order parameter (see e.g.\ \cite{Callaghan2001} and
references therein). In these systems, the non-trivial rheological
response is closely related to a variation of the system structure
accompanied by a change in the nematic order parameter. However,
such an order parameter-related structural change is absent
in the case of our glass forming model.}

Nevertheless, a qualitative similarity to the case of present
simulations may be found when red blood cells are concerned. This
similarity rests upon the fact that profile bunting in red blood
cells occurs only if a certain amount of aggregation among red blood
cells is present (see e.g.\ Fig.\ 7 in reference
\cite{Bishop2001a}). The aggregation enhances the solid character of
the suspension and leads to a higher yield stress, similar to a
reduction of temperature in our model.

The width of the ``jammed'' region can be estimated
from a knowledge of the yield stress in the system via $\sigma(z=W/2) =
\rho\Fe W/2 = \sigmay$ which gives  $W = 2\sigmay/(\rho\Fe)$.
The question arises whether the static or the dynamic yield stress
should be used for an estimate of $W$. Results of our simulations
suggest that the dynamic yield stress yields a better
estimate of the width of the solid-like region in the channel
center. This is exemplified in \fref{fig:yieldstressLJ}. In this figure,
two vertical dashed lines mark the bounds for the solid-like region
estimated via dynamic yield stress whereas the bounds denoted
by short vertical solid lines are obtained using the static yield stress.
As a survey of the velocity gradient reveals, the use of $\sigmaystatic$
overestimates the width of the solid-like region.

\blue{It is interesting to check the origin of the observed finite shear rates at stresses above
the dynamic but below the static yield stress.} For this purpose, we performed a
series of long simulation runs of a smaller system ($L_x \times L_y \times L_z=10 \times 10 \times 40$)
at constant imposed stress for a temperature of $T=0.2$ (far below $\Tc$).
All simulations started after an initial aging of the system during a time of
$\tw=10^4$. \blue{Interestingly, our data reveal the presence of a
stick-slip like plastic deformation. This occurs not only at stresses between
$\sigmaydynamic$ and $\sigmaystatic$ but also at stresses \emph{below} $\sigmaydynamic$.}

\begin{figure}
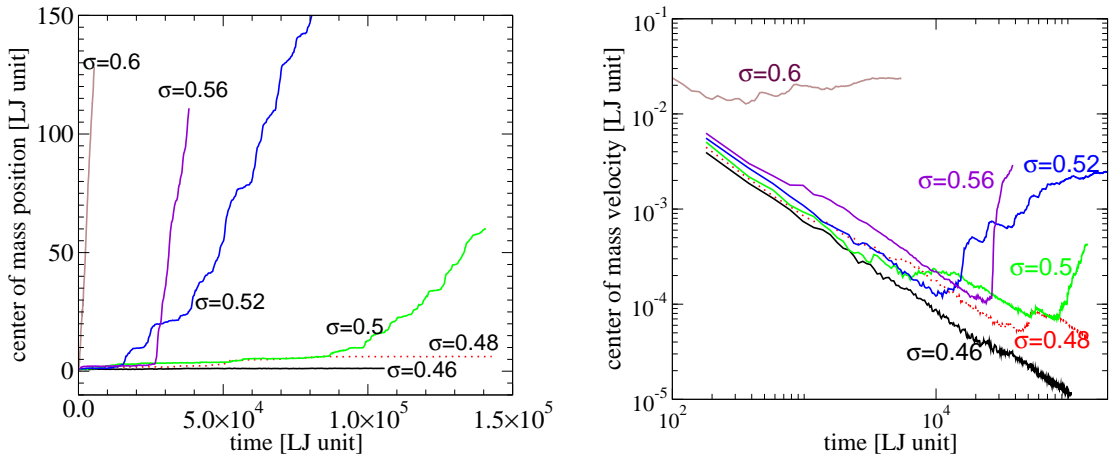

\epsfig{file=fig4a.ps,height=60mm,clip=}\hspace*{5mm}
\epsfig{file=fig4b.ps,height=60mm,clip=}
\caption[]{(Color online) Left: $X$-position of the center of mass of the fluid
 versus time for some selected stresses.
The results shown here correspond to simulations under constant
imposed stress of the present binary LJ model at a temperature of
$T=0.2$ (below $\Tc$). The dimensions of the simulation box are $L_x
\times L_y \times L_z=10 \times 10 \times 40$. Right: The same data
as in the left panel, now divided by time. Note the presence of a
flow with a time-independent velocity at $\sigma=0.6$. All
quantities are in LJ units.} \label{fig:creep}
\end{figure}

This behavior is illustrated in the left panel of \fref{fig:creep}.
The panel shows the center of mass position of the whole
fluid versus time for some selected values of the imposed stress ranging
from $\sigma=0.46$ (below the dynamic yield stress) up to the static yield
stress ($\sigma=\sigmaystatic=0.6$). The corresponding center of mass velocity
is depicted in the right panel of the same figure. As seen from this panel,
for stresses below $\sigma=0.52$, the contribution of the stick-slip like motion
to the flow velocity drops by roughly an order of magnitude. \blue{It is noteworthy
that a qualitatively similar stick-slip like dynamics has also
been observed in our previous simulations under imposed \emph{shear} \cite{Varnik2003}
as contrasted to the present case of imposed \emph{stress}.}

Using the data shown in \fref{fig:creep}, we can estimate the
contribution of this intermittent motion to the overall shear rate via
the relation $\bar{\gammadot}=V_{\mathrm{cm}}/L_z$. Using $L_z=40$, this gives
$\bar{\gammadot}=
5.9   \times 10^{-4},\;
7.3 \times 10^{-5}, \;
6.1 \times 10^{-5}, \;
1.1 \times 10^{-5}, \;
1.1 \times 10^{-5}$ and $2.8 \times 10^{-7}$ for $\sigma=0.60, \;0.56, \;0.52, \;0.50, \;0.48$
and $0.46$ respectively. The finite shear rate observed in the case of stresses
between $\sigma=0.5$ and $\sigma=0.6$ is therefore closely related to
the onset of significant stick-slip motion at these stresses.

One interesting consequence of the presence of a yield stress is that,
for a given driving force (pressure gradient) a flow blockage may occur
if the channel width is too small to allow the formation of stresses above the
system's yield stress. This 'critical' channel width is simply
estimated via $L_z = 2 \sigmay/(\rho\Fe)$. Obviously, a similar situation
would also occur at constant channel width via decreasing the applied body
force below $\Fe = 2\sigmay/(\rho L_z)$.  Let us estimate this 'threshold'
$\Fe$. At the temperature and density studied here ($T=0.2$
and $\rho=1.2$) the system exhibits a dynamic yield stress of  $\sigmaydynamic\approx 0.5$.
Using this value as the yield stress along with $L_z=80$ we obtain
$\Fe \approx 0.01$ for the minimum force per particle required to induce
a flow in the system.

\begin{figure}
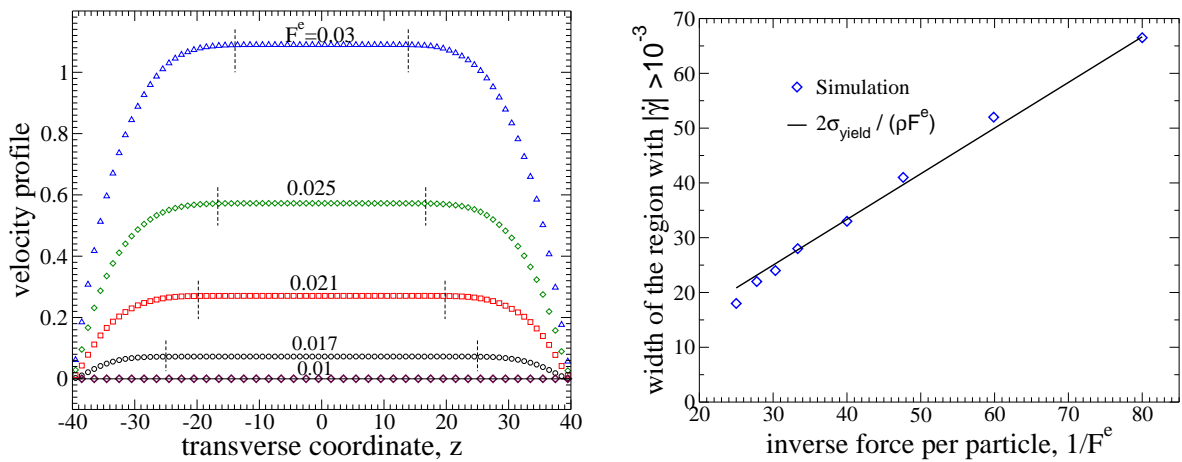

\epsfig{file=fig5a.ps,height=60mm,clip=}\hspace*{5mm}
\epsfig{file=fig5b.ps,height=60mm,clip=} \caption{(Color online)
Left: Velocity profiles (LJ units) obtained for various choices of
the external force per particle as indicated. At the temperature and
density studied here ($T=0.2$ and $\rho=1.2$) the system exhibits a
dynamic yield stress of  $\sigmay\approx 0.5$. Using this value as
the threshold stress for the onset of shear flow along with the
relation $\sigma(z) = \rho \Fe z$ for the local stress in the
channel, the expected width of a central region with a solid-like
behavior (flat velocity profile) can be estimated via
$W=2\sigmay/(\rho\Fe)$. The short vertical dashed lines help to
illustrate this central region. Beyond this part of the channel, a
liquid-like behavior is expected. As the external body force reaches
a value of $\Fe \approx 0.01$, the 'jammed' region comprises the
whole channel. Right: The width of the solid-like region is
estimated approximately as the width of a region with a shear rate
larger than $\gammadot=10^{-3}$ (see the left panel of
\fref{fig:Tprofile} for an inspection of shear rates). The solid
line gives the above mentioned expected linear dependence on
$1/\Fe$. Deviations at high $\Fe$ are probably related to undesired
viscous heating which leads to an enhanced softening of the solid
region (see also a discussion of the velocity gradients). All
quantities are given in LJ units. } \label{fig:vprofile}
\end{figure}

This aspect is illustrated in the left panel of
\fref{fig:vprofile}, where velocity profiles are shown for various
choices of $\Fe$. As expected, the width of the 'jammed' region
increases with decreasing $\Fe$. For $\Fe=0.017$, for example, there
remains a narrow liquid-like region close to the walls while the rest
of the system is in a 'jammed' state. This observation is made more
quantitative in the right panel of the same figure.
For this purpose we determine for each $\Fe$ the width of a region
with a shear rate larger than a small but finite value.
We find that the choice $|\gammadot| = 10^{-3}$ is a good
choice for an accurate determination of the position of
the 'interface' between the liquid-like and the solid-like regions.
As can be seen from the right panel of \fref{fig:vprofile},
results of this analysis obey well the expected
relation $W=2\sigmay/(\rho\Fe)$ if $\sigmay = \sigmaydynamic \approx 0.5$
is used. The use of static yield stress ($\sigmaystatic\approx 0.6$)
would overemphasize the size of the solid-like part of the channel.

\begin{figure}
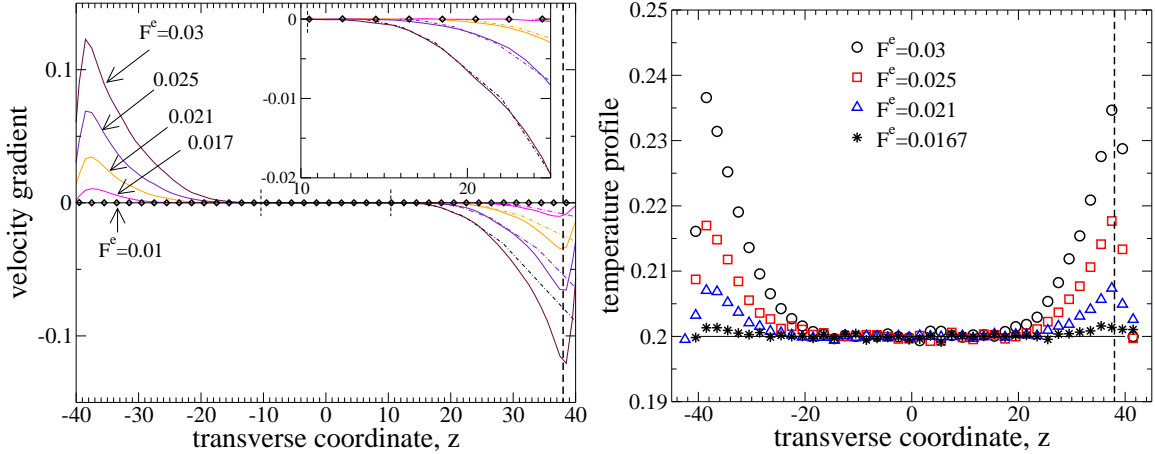

\epsfig{file=fig6a.ps,height=60mm,clip=}
\epsfig{file=fig6b.ps,height=60mm,clip=} \caption{(Color online)
Left: Shear rate, $\gammadot(z)=\partial u(z)/\partial z$, for
exactly the same choices of $\Fe$ as in the left panel of
\fref{fig:vprofile}. As expected, the width of the central
solid-like region ($\gammadot \approx 0$) increases upon decreasing
the external force per particle and eventually reaches the width of
the channel at $\Fe \approx 0.01$ (see also the left panel of
\fref{fig:vprofile}). The dotted dashed lines to the right of the
panel represent velocity gradients computed using the flow curve
($\sigma-\gammadot$) of the same model under homogeneous shear
\cite{Varnik2006} and the fact that $\sigma(z)=\rho\Fe z$ (i.e., by
plotting for each $z$ the shear rate $\gammadot(\sigma(z))$). As
seen from the inset, for $z\le 20$ the thus obtained $\gammadot(z)$
agrees well with the observed velocity gradient. However,
significant deviations occur close to the walls (main figure). These
deviations as well as the maxima in $|\partial u(z)/\partial z|$ are
probably related to a local increase of temperature. A vertical
dashed line at $z=38$ marks the approximate position of these
maxima. As a survey of the right panel reveals, temperature profiles
also exhibit maxima at approximately the same transverse position.
Right: Temperature profile across the channel for various choices of
the external force per particle as indicated. The temperature
profiles develop maxima in the proximity of the walls. The magnitude
of the excess temperature with respect to the prescribed value
($T=0.2$ in the present case) increases at higher $\Fe$. In addition
to a stress-induced decrease of shear viscosity (shear thinning)
this temperature increase enhances the decrease of the local
viscosity further. As a result, the local shear rate,
$\gammadot(z)$, increases faster than would be expected on the basis
of shear thinning alone. As in the case of the left panel, a
vertical dashed line at $z=38$ marks the approximate position of
temperature maxima. All quantities are given in LJ units. }
\label{fig:Tprofile}
\end{figure}

The left panel of \fref{fig:Tprofile}
illustrates the velocity gradients, $\partial u(z)/\partial z$,
for exactly the same values of the external force per particle as
shown in the left panel of \fref{fig:vprofile}. Here,
 velocity gradients are compared with profiles of shear rates,
$\gammadot(\sigma(z))$, estimated from a knowledge of the local stress
and the flow curve upon homogeneous shear: For each $z$,
the corresponding stress is first evaluated via $\sigma=\rho\Fe z$.
\blue{In order to estimate the shear rate corresponding to this shear stress,
we use the homogeneous flow curves ($\gammadot-\sigma$) determined
in our previous simulations \cite{Varnik2006}. In  these simulations,
an algorithm capable of ensuring a constant shear rate across the system was
used (see e.g.\ \cite{Varnik2006b} for details of the simulation method).}

As the inset of the left panel  of \fref{fig:Tprofile}
demonstrates, $\partial u(z)/ \partial z$ and $\gammadot(\sigma(z))$
agree well within a certain region in the channel, whose extension
increases upon decreasing the externally imposed force, $\Fe$.
This trend is probably related to the variation of the temperature
across the system.

Indeed, the right panel of \fref{fig:Tprofile} shows that
the system temperature is not constant overall in the channel but
slightly deviates from the prescribed value in a region between the
channel center and the walls. The extent of the central region with a constant
temperature increases at lower $\Fe$ while at the same time the magnitude of
the excess temperature with respect to the prescribed value decreases.
Noting that significant temperature excesses occur only as the shear stress
reaches values comparable to twice the dynamic yield stress of the model,
it is not surprising that, at such high stresses, even a local thermostat is
not able to ensure a constant temperature profile across the system.

\begin{figure}
\epsfig{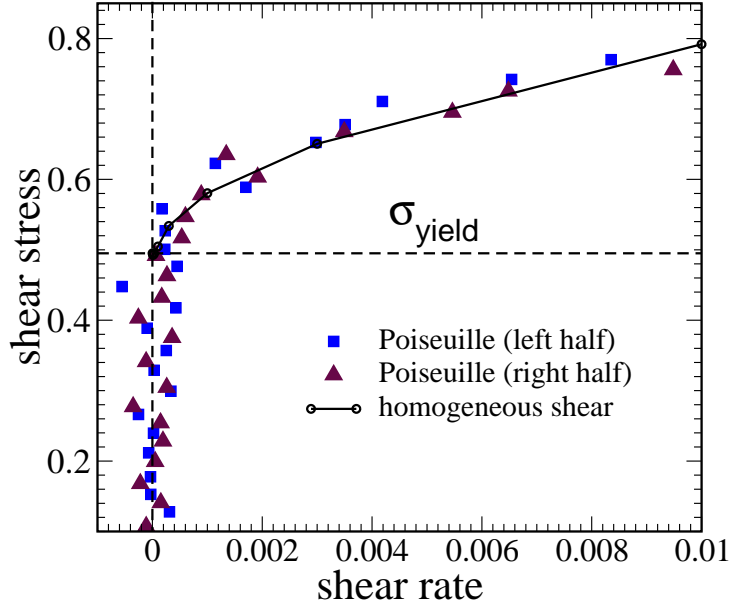} \caption{(Color online)
Shear stress versus shear rate. Filled symbols correspond to the
data obtained from an analysis of the relation between the local
shear rate and the local shear stress across the channel in the
present simulations ($\Fe=0.0167$). Squares correspond to the data
extracted from the left part of the channel and triangles to the
right half of the simulation box. Connected circles reproduce the
flow curve obtained from previous simulations of a homogeneous shear
for the same model and at the same density and temperature as
studied here ($T=0.2$, $\rho=1.2$) \cite{Varnik2006}. A horizontal
dashed line marks the position of the stress plateau which, for the
present purpose, plays the role of a dynamic yield stress,
$\sigmaydynamic \approx 0.5$ (this stress plateau becomes visible if
the data are plotted in log-scale as shown in \cite{Varnik2006}. On
a linear plot, data points for shear rates below, say
$\gammadot=10^{-4}$ are not distinguishable). For shear stresses
below $\sigmaydynamic \approx 0.5$ the shear rates obtained from the
present Poiseuille-type flow are scattered around  $\gammadot=0$
indicating that the shear rate vanishes for $\sigma\le
\sigmaydynamic$. All quantities are given in LJ units.}
\label{fig:flow-curves}
\end{figure}

In the above, the knowledge of the homogeneous flow curve is used in order
to obtain an independent estimate of the local shear rate across the channel.
Similarly, one can use the fact that the shear stress is well known in the
channel, $\sigma(z)=\rho\Fe z$, along with the knowledge of the velocity
gradient in order to obtain the flow curves, i.e.\ shear stress as a function
of shear rate (velocity gradient). For this purpose, we plot in
\fref{fig:flow-curves} for each value of $z$ the corresponding
values of $\sigma(z)$ versus $\gammadot(z)$.
Due to the symmetry around the mid-plane of the channel, the above procedure
would yield identical flow curves using the data from the left and right
halves of the channel provided that no statistical uncertainty is present.
In reality, however, there is a finite statistical scatter. In order to
illustrate this fact, we do not average the results but depict the individual
flow curves obtained from the analysis of each half of the channel.
The so obtained $\sigma-\gammadot$ curves are then compared to the result of
simulations at homogeneous shear \cite{Varnik2006}. The comparison is
restricted to shear stresses, where the relative deviations between the local
temperature and the prescribed one are below one percent.
With this restriction, the flow curve obtained from the present simulations
agree well with that of homogeneous shear.

\begin{figure}
\epsfig{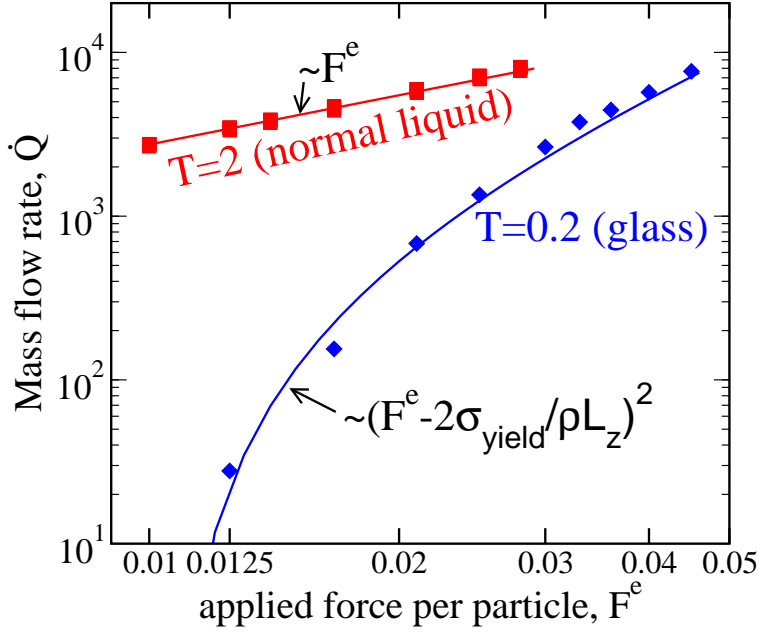} \caption{(Color online) Mass
flow rate, $\Qdot= 2\rho L_y\int_0^{Lz/2}u(z)dz$, versus applied
force per particle for two temperatures characteristic of the normal
liquid state ($T=2$) and the glassy phase ($T=0.2$). In the normal
liquid state, one expects $u(z) \propto \Fe$ (more precisely,
$u(z)=\Fe\rho (L_z^2/4-z^2)/(2\eta)$) and hence $\Qdot \propto \Fe$.
In the glass, on the other hand, the velocity profile is blunted due
to the presence of a yield stress. We expect the system to stop
flowing for sufficiently weak external forces, $\Fe<2\sigmay/(\rho
L_z)(\approx 0.01$ for the case studied here). The simulation
results shown here conform this picture. Please note that, while the
relation $\Qdot \propto \Fe$ is an exact result for the normal
liquid state, the fit $\Qdot \propto (\Fe-2\sigmay/(\rho L_z))^2$ is
a purely empirical one motivated by the fact that $\Qdot$ is
expected to vanish at $\Fe = 2 \sigmay / (\rho L_z)$. All quantities
are given in LJ units.} \label{fig:debit}
\end{figure}

Finally, \fref{fig:debit} illustrates how the mass flow rate,
$\Qdot= 2\rho L_y\int_0^{Lz/2}u(z)dz$, varies with applied force
per particle. As a survey of the velocity profiles (\fref{fig:vprofile})
already suggests, the mass flow rate rapidly decreases as the 'critical'
force per particle $\Fe\approx 0.01$ is approached
(note the logarithmic scale for the $y$-axis in \fref{fig:debit}).
In order to highlight the effect of a finite yield stress,
the data corresponding to the same model in the normal liquid state
(where the yield stress is identically zero)
are also shown. For the normal liquid state,  it is straightforward to show
that $\Qdot\propto \Fe$. This relation is nicely born out by
by the simulated data. We are, however, not aware of an analytic
expression for the mass flow rate in a yield stress fluid.
Therefore, we fit the data to the simplest polynomial in
$\Fe-2\sigmay/(\rho L_z)$ which describes the
simulated data best.

\section{Summary}
\label{sec:summary}
In this paper, we report on the flow behavior of a simple glass forming system.
The model consists of a 80:20 binary Lennard-Jones mixture first introduced
by Kob and Andersen in the context of the dynamics of supercooled systems
\cite{Kob1994,Kob1995,Kob1995a}. It is well known for its capability to form
a disordered solid at low temperatures/high densities \cite{Kob1997}.

Previous studies of the rheological response of the model
suggest the existence of a finite yield stress, $\sigmay$, in the glassy phase.
In particular, at a temperature of $T=0.2$ (deep in the glassy phase),
a finite static yield stress of $\sigmaystatic\approx 0.6$
was found via simulations upon imposed stress \cite{Varnik2004}. More recent studies
of the same model under imposed shear, on the other hand, showed the existence
of a stress plateau in the low shear rate limit of the flow curve
(stress versus shear rate) in the glassy phase \cite{Varnik2006}.
As the present work also underlines, this stress plateau paly
the role of a (dynamic) yield stress.

Here, we study a consequence of the presence of a finite yield stress
as a flow is induced in a planar channel via the application
of an external force. The stress in such a Poiseuille-type flow is
a linear function of the distance from the channel center. A two phase
behavior may, therefore, occur in the glassy phase
provided that the channel width is sufficiently
large in order to ensure that stress close to the walls (where it reaches
its maximum) is higher than the yield stress of the system.
While the system response is expected to be solid-like
 (zero shear rate) in a central part of the channel
(defined as a region where the stress is below
the system's yield stress), it should flow in the 'wings'
delimited by this central solid-like region and the walls.
This expectation is born out by our simulations (\fref{fig:yieldstressLJ}).

Furthermore, using the velocity gradient across the channel (\fref{fig:Tprofile}),
we define the width of the 'jammed' phase as the size
of the region with a shear rate of $\gammadot(z) = \partial u(z)/\partial z <10^{-3}$.
The accuracy of this estimate is demonstrated in the right panel of
\fref{fig:vprofile}, where it is shown that the relation
$W=2\sigmay/(\rho\Fe)$ is well satisfied by the data obtained from
the above analysis. As to the
numerical value of $\sigmay$ used in the above formula,
our simulation results are consistent with a value of
$\sigmay = \sigmaydynamic \approx 0.5$ (=stress plateau upon imposed shear \cite{Varnik2006})
while the use of $\sigmaystatic=0.6$ \cite{Varnik2004}
seems to overemphasize the size of the solid-like region.
Our simulation results also clearly show that, as  the stress increases
above the dynamic yield stress, the contribution of a stick-slip like motion to the
overall shear rate increases significantly (\fref{fig:creep}).

As a consistency check, flow curves obtained from the present Poiseuille-type
simulations are compared to the result of previous simulations under
homogeneous shear \cite{Varnik2006}. With the exception of relatively high
driving forces where uncontrollable viscous heat significantly biases
the present simulation results in the the vicinity of the walls,
good agreement is found between both approaches
(Figs.\ \ref{fig:Tprofile} and \ref{fig:flow-curves}).

Finally, the dependence of the mass flow rate on the externally imposed
force is studied. It is shown that the flow fully stops as
$\Fe \le 2\sigmay/(\rho L_z)$. Again, simulated data are well described by
this formula provided that $\sigmay=\sigmaydynamic \approx 0.5$
is used. In particular, the use of $\sigmaystatic=0.6$ leads to $\Fe=0.0125$
for a complete arrest of the flow, in contrast with our simulations
where a finite flow is observed for this value of the external force per
particle (\fref{fig:debit}).

It is interesting to note that similar (blunted) shapes of the velocity
profile are also observed in pressure driven flows of both
neutrally buoyant suspensions of spheres (with a size of the order of 1mm)
\cite{Hampton1997,Han1999} as well as red blood cells
(biconcave disks of $2\mu$m thickness and $8\mu$m diameter)
\cite{Tangelder1986,Bishop2001a}. In these cases, however, the profile
blunting is usually accompanied by a migration of particles from the
wall region towards the center of the channel ('wall migration') a phenomenon,
which is absent in the case of present studies (see the right panel
of \fref{fig:yieldstressLJ}).

Nevertheless, a qualitative similarity to the case of present simulations
may be found when red blood cells are concerned. This similarity
rests upon the fact that profile bunting in red blood cells occurs only if a
certain amount of aggregation among red blood cells is present (see e.g.\
Fig.\ 7 in reference \cite{Bishop2001a}). The aggregation gives rise to a
finite yield stress, an important feature whose effects on the flow
profile are the focus of the present work.

\section*{Acknowledgments}
We thank Jean-Louis Barrat, Lyderic Bocquet for fruitful discussions
who strongly motivated the present studies. Parts of the MD simulations related to this
work have been performed during the stay of F.V. in
Laboratoire de Physique de la Mati\`ere Condens\'ee et Nanostructure (LPMCN),
Universit\'e Claude Bernard, Lyon 1. The financial support by the
Max-Planck-Initiative ``Multiscale Materials Modelling of Condensed Matter'' (MMM)
and by the DFG Priority Programme Nano-\& Microfluidics (SPP1164, project Va 205/3-2)
during the preparation of this manuscript is also acknowledged.

\end{document}